\documentstyle[11pt]{article}
\newlength{\extraspace}
\newlength{\extraspaces}
\newcommand{\be}{\begin{equation}
\addtolength{\abovedisplayskip}{\extraspaces}
\addtolength{\belowdisplayskip}{\extraspaces}
\addtolength{\abovedisplayshortskip}{\extraspace}
\addtolength{\belowdisplayshortskip}{\extraspace}}
\newcommand{\ee}{\end{equation}}

\newcommand{\ba}{\begin{eqnarray}
\addtolength{\abovedisplayskip}{\extraspaces}
\addtolength{\belowdisplayskip}{\extraspaces}
\addtolength{\abovedisplayshortskip}{\extraspace}
\addtolength{\belowdisplayshortskip}{\extraspace}}
\newcommand{\ea}{\end{eqnarray}}

\newcommand{\R}{\bf R}

\begin{document}
\begin{center}
{\Large\bf On the Singularities of Reissner-Nordstr\"{o}m
Space-Time}
\end{center}
\bigskip

\centerline{\bf M. Abdel-Megied and Ragab M. Gad}
\bigskip
\centerline{{\bf Mathematics Department, Faculty of Science}}
\hspace{2.6cm}{\bf Minia University, 61915 El-Minia, Egypt}
\footnote[1]{e-mail:amegied@frcu.eun.eg, ragab2gad@hotmail.com}
\hspace{1cm}
\\
\\
\section*{{\bf Abstract}}

It is shown that if two Reissner-Nordstr\"{o}m space-times, both
with the same mass m and charge e, glued together in the singularities, then the light ray in black hole of the first space-time can go continuously through the
singularity into black hole of the second. The behavior of tidal
forces near the Reissner-Nordstr\"{o}m space-time singularity   is
examined by considering what happens between two particles falling
freely towards the singularity.



\section{{\bf Introduction}}
   A well known spherically symmetric solution of the Einstein Maxwell
equations is the Reissner-Nordstr\"{o}m space-time. It represents
a non-rotating black hole with a mass m and a charge e. The basic
properties of this solution are discussed by many authors
\cite{2}, \cite{10}, and \cite{13}.
\par
   Gad \cite{4} studied the possibility of gluing two Schwarzschild space-times in
their singularities. It is shown that if two Schwarzschild
space-times, both with the same mass, glued together in the
singularities, then the light ray in black hole of the first
space-time can go continuously through the singularity into black
hole of the second. This study was by well defining of the passage
of particles through the singularity. Although such gluing of
singular points is, of course, not possible within the theory of
differential manifolds, because the differential manifold
structure of space-time breaks down at a singularity.
\par
   In the last two decades , somome mathematical frameworks have been found to generalize the
space-time model and consider it to be a differential space rather
than a differential manifold. For instance, a generalized model of
space-time has been studied in a series of papers by Gruszczak
\cite{7} - \cite{9}, and Heller et al \cite{11}. The theory of
differential spaces open the possibilities to include
singularities of space-time.  We  use here a special form of
differential spaces studied in \cite{5}, \cite{6}, \cite{GB96}.
\par
   The aim of this paper is twofold, one is to study the possibility of gluing two
Reissner-Nordstr\"{o}m space-times in their singularities, the
second is to study the behavior of tidal forces near
Reissner-Nordstr\"{o}m space-time singularity by examining
what happens between two particles falling freely towards the singularity.\\
In the this work, the Reissner-Nordstr\"{o}m solution together
with its singularities is consider as d-space \cite{5}, \cite{6}.
\\
\par
 Since the theory of differential spaces is not commonly known,
 we give its basic definition in sec. \ref{sec2}.
  A brief discussion of the geodesics in the
Reissner-Nordstr\"{o}m space-time is given in sec. \ref{sec3}. In
sec. \ref{sec4} it is shown that: in the case of radial null
geodesics, two Reissner-Nordstr\"{o}m space-times can be glued in
their singularities. The behavior of the Jacobi vector fields near
the Reissner-Nordstr\"{o}m singularity are studied in sec.
\ref{sec5}.

\section{\bf{d-Spaces}}\label{sec2}

Since the theory of d-spaces, introduced by \cite{5}, \cite{6}
\cite{GB96}, is not widely known, we will present in this section
the necessary basic concepts of d-spaces.

{\bf{Definition 1:}}  Let $M$ be a topological space. The pair
$(M,C)$ is called {\bf{differential space}} ( or {\bf{d-space}}),
and $C$ {\bf{differential structure}}, if $C$ is a sheaf of
continuous real-valued functions on $M$ which form an algebra
(w.r.t. pointwise operation). This definition generalizes
Sikorski's definition \cite{S71} which requires global functions.
In addition, the composition of function $g \in
C^{\infty}(\R^{n})$ with $n$ arbitrary functions of $C$ must again
be a function of $C$, and $M$ must carry the initial topology of
$C$. The latter axiom is one of the reasons, why Sikorski's
definition is not able to describe the singularities in general
relativity \cite{HS94}. Therefore, Heller and Sasin proposed
\cite{HS95} to use Mostow spaces \cite{M79} which  called later
structured spaces. These
spaces are slightly more special than the d-spaces of definition 1.\\

{\bf {Definition 2:}}  Let $(M,C)$ be a d-space, $x \in M$,
$C_{x}$ the stalk at $x$. A map
$$
V :C_{x} \rightarrow
$$
is called {\bf{tangent vector}} to $(M,C)$ in $x$, if for all $n
\in Natu$, all $f_{1},..., f_{n} \in C_{x}$, and all germs
$\alpha$ of $C^{1}(\R^n)$ at $y := (f_{1}(x), ..., f_{n}(x)) \in
\R^n$, the equation
$$
 V(\alpha \circ (f_{1}(x), ..., f_{n}(x))) = \Sigma_{i=1}^{n} (\partial_{i}\alpha) \cdot V(f_{i})
$$
holds, provided $\alpha \circ (f_{1}(x), ..., f_{n}(x))\in C_{x}$.
Here $\partial_{i}\alpha$ denotes the partial derivative
of $\alpha$ w.r.t. the $i$-th argument.\\
The vector space of all tangent vectors to $(M,C)$ in $x$ is
called {\bf{tangent space}} $T_{x}M$.

The dimension of $T_{x}M$ can easily be computed:\\
We call a subset $F$ of the stalk $C_{x}$ {\bf{dependent}}, if
there is a germ $g \in F$ and a finite subset $\{f_{1}, ...,
f_{n}\} \subseteq F$ not containing $g$ such that
$$
 g = \alpha \circ (f_{1}, ..., f_{n})
$$
holds for some $C^{1}$-functions $\alpha$. Now choose any
independent subset $G \subseteq C_{x}$ such that all germs $f \in
C_{x}$ are a combination of finitely many elements of $G$, i.e.
\begin{equation}
f = \alpha \circ (f_{1}, ..., f_{n})
\end{equation}
holds for some $n$, some local $C^{1}$-function $\alpha$ on
$\R^{n}$, and some $f_{1}, ..., f_{n} \in G$. Write $G$ as $G = \{
f_{j}\}_{j\in J}$. Then definition 2 shows that $\{V_{j}\}_{j \in
J}$ defined by
$$
V_{j}(f_{k}) = \delta_{jk}
$$
is a basis of $T_{x}M$.\\
On d-spaces, differentiation, differential forms and exterior
derivative are introduced \cite{5}, \cite{6} in a way similar to
Sikorski's differential spaces.
\par
The topology of space-times with singularities has been a subject
of frequent discussions (e.g. in \cite{B76}, \cite{G68},
\cite{HS95}). Also the definition of a d-space requires an a
priory given topology. But definition 2 shows that the initial
topology of $C$ cannot be finer than this a priory topology. This
leads to the following definition {\bf {Definition3:}} Let $(M,
\varrho)$ be a topological space, and $F$ a sheaf of local
function $M$ defined w.r.t. the topology $\varrho$. A topology
$\alpha$ on $M$ that is coarser than $\varrho$, is called a
{\bf{slackening}} of $F$, if for every $V \in \varrho$ and every
$f \in F (V)$, there are $U\in \alpha$; $U\supseteq V$, and $g \in
F (U)$ such that$f = g|_{V}$ holds. It is shown in \cite{5},
\cite{6} that there exits a coarsest slackening $\mu$, such that
all functions $g \in F(U)$; $U\in\mu$
are continuous. This $\mu$ is called {\bf{initial topology}} of $(M,C)$\\
It is not difficult to construct this initial topology explicitly
(see \cite{5}, \cite{6}).

\section{\bf{Geodesics in the Reissner-Nordstr\"{o}m Space-Times}}\label{sec3}
\par
The Reissner-Nordstr\"{o}m solution is given by the following
metric \cite{10}
\begin{equation}\label{1.1}
ds^2 = - (1 - \frac{2m}{r} + \frac{e^2}{r^2}) dt^2 + (1 -
\frac{2m}{r} + \frac{e^2}{r^2})^{-1} dr^2 + r^2 (d\theta^2 +
\sin^{2}\theta d\phi^2),
\end{equation}
where m represents the gravitational mass and e the electric
charge of the body.
\par
   The geodesic equations for the metric (\ref{1.1}) are given by the following equations
 $$
\frac{du^1}{d\tau} + \frac{e^2 - mr}{r(r^2 - 2mr + e^2)} (u^1)^2 -
\frac{r^2 - 2mr + e^2}{r} (u^2)^2 - \frac{r^2 - 2mr + e^2}{r}
\sin^{2}\theta (u^3)^2 +
$$
\begin{equation} \label{2.1}
\frac{(\frac{m}{r^2} - \frac{e^2}{r^3})(r^2 - 2mr + e^2)}{r^2}
(u^4)^2 = 0,
\end{equation}

\begin{equation}\label{2.2}
\frac{du^2}{d\tau} + \frac{2}{r}u^1 u^2 - \sin\theta\cos\theta
(u^3)^2 = 0,
\end{equation}
\begin{equation}\label{2.3}
\frac{du^3}{d\tau} + \frac{2}{r}u^1 u^3 + 2\cot\theta u^2 u^3 = 0,
\end{equation}
\begin{equation}\label{2.4}
\frac{du^4}{d\tau} + \frac{2( - m + \frac{e^2}{r})}{r^2 - 2mr +
e^2}u^1 u^4 = 0,
\end{equation}
where $\tau$  is an affine parameter, $u^a = \frac{dx^a}{d\tau}$
and $x^a(\tau)$ are
the coordinates of a space-time point on the geodesic.\\
Let the geodesic $\gamma$ be given by $\gamma (\tau) = (r(\tau),
\theta(\tau), \phi(\tau), t(\tau))$. Without loss of generality We
 suppose that $\gamma(\tau_{0}) = (r_{0}, \theta_{0},
\frac{\pi}{2}, t_{0})$ and for all $\tau \, : \phi =
\frac{\pi}{2}$. Then $\frac{d\phi}{d\tau} = u^3 = 0$  and the
above equations become
 $$
\frac{du^1}{d\tau} + \frac{e^2 - mr}{r(r^2 - 2mr + e^2)} (u^1)^2 -
\frac{r^2 - 2mr + e^2}{r} (u^2)^2  +
$$
\begin{equation} \label{2.5}
\frac{(\frac{m}{r^2} - \frac{e^2}{r^3})(r^2 - 2mr + e^2)}{r^2}
(u^4)^2 = 0,
\end{equation}
\begin{equation}\label{2.6}
\frac{du^2}{d\tau} + \frac{2}{r}u^1 u^2  = 0,
\end{equation}
\begin{equation}\label{2.7}
\frac{du^3}{d\tau}  = 0,
\end{equation}
\begin{equation}\label{2.8}
\frac{du^4}{d\tau} + \frac{2( - m + \frac{e^2}{r})}{r^2 - 2mr +
e^2}u^1 u^4 = 0.
\end{equation}
Integrating equations (\ref{2.6}) and (\ref{2.8}), we obtain
respectively
\begin{equation}\label{2.9}
u^2 = \frac{c_{2}}{r^2},
\end{equation}
\begin{equation}\label{2.10}
u^4 = \frac{c_{1}r^2}{r^2 - 2mr + e^2},
\end{equation}
where the integrating constant $c_{1}$  represents the energy $E$
(at $r \rightarrow \infty$ ) of a test
particle and $c_{2}$  the angular momentum $L$  \cite{13}.\\
Substituting (\ref{2.9}) and (\ref{2.10}) in (\ref{1.1}), using
the condition $u^3 = 0$ , we get
\begin{equation}\label{2.11}
(u^1)^2 = c_{1}^{2} + (k - \frac{c_{2}^{2}}{r^2})( 1 -
\frac{2m}{r} + \frac{e^2}{r^2}).
\end{equation}
For timelike geodesics, $k = - 1$, equation (\ref{2.11}) implies
 \begin{equation}\label{2.12}
u^1 = \pm \big[c_{1}^{2} - (1 + \frac{c_{2}^{2}}{r^2})( 1 -
\frac{2m}{r} + \frac{e^2}{r^2})\big]^{\frac{1}{2}}.
\end{equation}
Similarly, $k = 0$  corresponds to null geodesics, thus we have
 \begin{equation}\label{2.13}
u^1 = \pm \big[c_{1}^{2}  - (\frac{c_{2}^{2}}{r^2})( 1 -
\frac{2m}{r} + \frac{e^2}{r^2})\big]^{\frac{1}{2}}.
\end{equation}
The $\pm$ sign is used in equations (\ref{2.12}) and (\ref{2.13})
since $\frac{dr}{d\tau} > 0$  holds at each point on a geodesic
where it is outward and $\frac{dr}{d\tau} < 0$   where it is
inward.
\par
To investigate what happens when the null geodesics approach the
singularity $r = 0$. Therefore we shall be interested in geodesics
on which $r$ decreases as the affine parameter $\tau$ increases,
that is, $\frac{dr}{d\tau} < 0$. Hence the minus sign will be
chosen in equation (\ref{2.12}) and (\ref{2.13}). Also we restrict
ourselves to the region $0 < r < r_{-}$ (assuming $m >abs{e}$).
\section{Gluing of Two Reissner-Nordstr\"{o}m Space-Times in Their Singularities}\label{sec4}
Now we study the gluing of two Reissner-Nordstr\"{o}m space-times
$M_{1}$ and $M_{2}$ in the singularity, $M_{1}$ and $M_{2}$ being
given respectively as follows
$$
ds_{1}^2 = - (1 - \frac{2m_{1}}{r} + \frac{e_{1}^2}{r^2}) dt^2 +
(1 - \frac{2m_{1}}{r} + \frac{e_{1}^2}{r^2})^{-1} dr^2 + r^2
(d\theta^2 + \sin^{2}\theta d\phi^2),
$$
$$
ds_{2}^2 = - (1 - \frac{2m_{2}}{r} + \frac{e_{2}^2}{r^2}) dt^2 +
(1 - \frac{2m_{2}}{r} + \frac{e_{2}^2}{r^2})^{-1} dr^2 + r^2
(d\theta^2 + \sin^{2}\theta d\phi^2).
$$
First we notice that we must have $m_{1} = m_{2}$ and $e_{1} =
e_{2}$ , if $M_{1}\cup M_{2}$ is a differential space, that is, a
solution of Einstein's equation. Although there is not yet a
general theory of partial differential equations on differential
spaces, Einstein equations can be treated in this special case,
where an open dense subset of the space is a differential manifold
\cite{1}. So $M_{1}\cup M_{2}$  is one single solution with the
same integration constants, that is, $m_{1} = m_{2}$ and $e_{1} =
e_{2}$.
\par
   We consider radial null geodesics with zero angular momentum ($c_{2} = L = 0$),
then it follows from (\ref{2.13}), (\ref{2.10}) and (\ref{2.9})
that
\begin{equation}\label{3.1}
u^1 = \pm E,
\end{equation}
\begin{equation}\label{3.2}
u^4 =  E(1 - \frac{2m}{r} + \frac{e^2}{r^2})^{-1},
\end{equation}
\begin{equation}\label{3.3}
u^2 = u^3 = 0.
\end{equation}
Dividing (\ref{3.1}) by (\ref{3.2}), we obtain
\begin{equation}\label{3.4}
\frac{dr}{dt} = \pm (1 - \frac{2m}{r} + \frac{e^2}{r^2}).
\end{equation}
The minus sign is used in the region  $r < r_{-}$($r_{\pm} = m \pm
\sqrt{m^2 - e^2}$ are the roots of the quadratic $ \Delta = r^2 -
2mr + e^2$),
because in this region $r$ decreases with increasing $\tau$, that is, $\frac{dr}{dt} < 0$.\\
Integrating (\ref{3.4}) gives
\begin{equation}\label{3.5}
t = - r_{\star} +{constant}
\end{equation}
where
$$
r_{\star} = r + m \ln abs{r^2 - 2m + e^2} + \frac{2m^2 -
e^2}{\sqrt{m^2 - e^2}} \ln abs{\frac{r - r_{+}}{r - r_{-}}}.
$$
Equation (\ref{3.5}) must be contrasted with the equation
\begin{equation}\label{3.6}
r = \pm E\tau +text{constant},
\end{equation}
which relates $r$ to the proper time $\tau$. The surface $r =
r_{+}$  is an event horizon in the same sense that $r = 2m$ is an
event horizon in Schwarzschild space-time. Also the surface $r =
r_{-}$ is a horizon \cite{2}. Equation (\ref{3.5}) and (\ref{3.6})
show that while the radial geodesic crosses the horizon $r =
r_{-}$  in within finite proper time without ever noticing it, it
takes an infinite coordinate time even to arrive at the horizon.
This is clear from equation (\ref{3.2}): When the geodesic
approaches the horizon $r = r_{-}$
the tangent in the $t$-direction tends to infinity (see, Fig. (1)).\\
\par
\unitlength 0.65mm \linethickness{0.4pt}
\begin{picture}(135.00,75.00)
\put(60.00,20.00){\vector(1,0){70.00}}
\put(60.67,20.00){\line(0,1){50.00}}
\put(60.67,70.00){\line(0,0){0.00}}
\put(105.00,20.00){\line(0,1){50.00}}
\bezier{308}(100.33,70.00)(95.00,28.00)(60.67,27.33)
\put(105.00,15.00){\makebox(0,0)[cc]{$r=r_{-}$}}
\put(60.67,15.00){\makebox(0,0)[cc]{r=0}}
\put(60.67,20.00){\vector(0,1){50.00}}
\put(60.33,20.00){\vector(0,1){49.67}}
\put(59.67,75.00){\makebox(0,0)[cc]{t}}
\put(135.00,20.00){\makebox(0,0)[cc]{r}}
\put(78.67,10.00){\makebox(0,0)[cc]{Fig. (1)}}
\end{picture}

At the singularity $r = 0$ the tangent component in the
$t$-direction tends to zero and in the $r$-direction remains
constant. It is shown from the definition of black hole that in
the region $r < r_{-}$ no light ray can remain at a constant value
of $r$ but must move inwards to fall in the singularity $r = 0$
and it can never escape to larger values of $r$  to communicate
with external light ray. This means that any light ray fallen in
the black hole can never leave this black hole again.

\par
\unitlength 0.65mm \linethickness{0.4pt}
\begin{picture}(150.00,124.67)
\put(150.00,80.00){\line(-1,0){70.00}}
\put(79.67,80.00){\line(-1,0){0.33}}
\put(79.67,40.33){\line(0,1){79.67}}
\put(75.00,120.00){\line(0,-1){79.67}}
\put(75.00,80.00){\line(-1,0){70.00}}
\put(100.00,40.00){\line(0,1){79.33}}
\put(130.00,120.00){\line(0,-1){80.00}}
\put(55.00,120.00){\line(0,-1){80.33}}
\put(25.00,120.00){\line(0,-1){80.00}}
\bezier{196}(97.00,115.00)(98.33,84.67)(79.67,85.00)
\bezier{180}(58.33,115.00)(56.67,88.33)(75.00,85.33)
\bezier{196}(97.33,96.67)(99.00,67.00)(79.67,65.00)
\bezier{188}(57.33,96.33)(56.00,68.67)(75.00,65.00)
\bezier{168}(104.67,115.00)(100.00,94.67)(109.00,75.33)
\bezier{180}(109.00,75.33)(130.67,65.00)(124.67,45.00)
\bezier{148}(128.00,52.00)(129.67,69.67)(119.33,85.67)
\bezier{136}(120.00,84.33)(113.67,91.00)(107.67,114.67)
\bezier{300}(141.00,117.67)(127.00,96.00)(134.33,47.67)
\bezier{296}(147.33,117.67)(130.00,78.33)(142.00,49.67)
\bezier{176}(52.00,115.00)(52.67,88.67)(41.67,75.33)
\bezier{156}(41.67,75.67)(27.33,69.00)(29.33,46.00)
\bezier{140}(26.67,52.67)(25.67,69.67)(35.33,85.00)
\bezier{136}(35.00,84.00)(40.33,90.33)(47.33,115.00)
\bezier{308}(12.67,118.00)(29.00,89.67)(17.33,46.67)
\bezier{304}(6.33,117.33)(24.00,84.67)(11.00,48.00)
\put(77.33,35.00){\makebox(0,0)[cc]{r=0}}
\put(100.00,35.00){\makebox(0,0)[cc]{$r=r_{-}$}}
\put(130.00,35.00){\makebox(0,0)[cc]{$r=r_{+}$}}
\put(55.00,35.00){\makebox(0,0)[cc]{$r=r_{-}$}}
\put(25.00,35.00){\makebox(0,0)[cc]{$r=r_{+}$}}
\put(138.33,122.67){\makebox(0,0)[cc]{$I_{1}$}}
\put(113.67,122.67){\makebox(0,0)[cc]{$II_{1}$}}
\put(90.00,122.67){\makebox(0,0)[cc]{$III_{1}$}}
\put(63.33,122.67){\makebox(0,0)[cc]{$III_{2}$}}
\put(40.33,122.67){\makebox(0,0)[cc]{$II_{2}$}}
\put(10.33,122.67){\makebox(0,0)[cc]{$I_{2}$}}
\put(55.00,21.33){\makebox(0,0)[cc]{Fig. (2)}}
\end{picture}\\

   Consider two Reissner-Nordstr\"{o}m space-times$M_{1}$  and $M_{2}$, both with the same mass and the same charge, and glued together in the singularities.\\
The question arises: Can the light leave the black hole of $M_{1}$ to fall into the black hole of $M_{2}$?\\
To answer this question, we assume that the two black holes are
represented by the two regions $III_{1}$ and $III_{2}$  (see, Fig. (2)), and we see what happens at $r = 0$.\\
\par

    Now in region $III_{1}$ the end point of the trajectory may be
considered as initial condition ($u^1 = + E, \, u^4 = 0$ ) in
region $III_{2}$ at $r = 0$. Then a light ray in region $III_{1}$
falling into the singularity can go continuously through the
singularity into region $III_{2}$ along the world line
$\gamma_{2}$ (see Fig. (3) and Fig. (4)).
\par
   We considered the radial null geodesics, with zero angular momentum,
running into the black hole. Some of them have a well-defined
limit at  $r = 0$ (see equations (\ref{3.1}) and (\ref{3.2})).
This open the possibility to ask, what happens to geodesics beyond
the singularity? and we try gluing two space-times in their
singularities. This gluing can be viewed  from the mathematical
point of view: since the metric coefficients depend only on $r$,
we can see that the metric and its first and second derivatives
agree on the two sides of the glued two space-times. This satisfy
the junction condition for gluing two space-times \cite{12}.\\

\unitlength 0.65mm \linethickness{0.4pt}
\begin{picture}(136.00,125.00)
\put(75.00,79.33){\vector(1,0){55.00}}
\put(75.00,79.33){\vector(-1,0){45.00}}
\put(75.00,40.00){\vector(0,1){80.00}}
\put(120.00,40.00){\line(0,1){79.67}}
\put(40.00,40.00){\line(0,1){79.00}}
\put(98.67,96.33){\vector(-4,-3){0.2}}
\bezier{124}(113.67,119.67)(112.00,105.33)(98.67,96.33)
\bezier{104}(98.33,96.33)(86.33,88.33)(74.67,88.33)
\put(96.00,71.00){\vector(1,-4){0.2}}
\bezier{136}(75.00,88.33)(92.00,87.67)(96.00,71.00)
\bezier{120}(96.00,71.00)(103.33,65.00)(107.00,44.33)
\put(54.67,98.67){\vector(4,-3){0.2}}
\bezier{104}(43.67,118.67)(43.67,106.00)(54.67,98.67)
\bezier{88}(54.67,98.67)(66.00,92.00)(75.00,92.00)
\put(53.00,75.00){\vector(-1,-2){0.2}}
\bezier{128}(75.33,91.67)(57.33,87.33)(53.00,75.00)
\bezier{108}(53.00,75.00)(41.33,66.33)(43.33,53.67)
\put(96.67,106.33){\makebox(0,0)[cc]{$\gamma_{1}$}}
\put(58.33,106.33){\makebox(0,0)[cc]{$\gamma_{2}$}}
\put(136.00,79.33){\makebox(0,0)[cc]{r}}
\put(75.00,125.00){\makebox(0,0)[cc]{t}}
\put(24.33,79.33){\makebox(0,0)[cc]{r}}
\put(75.00,35.67){\makebox(0,0)[cc]{r=0}}
\put(120.00,35.00){\makebox(0,0)[cc]{$r=r_{-}$}}
\put(40.00,35.00){\makebox(0,0)[cc]{$r=r_{-}$}}
\put(94.67,121.00){\makebox(0,0)[cc]{$III_{1}$}}
\put(58.00,121.33){\makebox(0,0)[cc]{$III_{2}$}}
\put(75.00,24.00){\makebox(0,0)[cc]{Fig. (3)}}
\end{picture}\\
\unitlength 0.65mm \linethickness{0.4pt}
\begin{picture}(140.33,114.00)
\put(75.00,70.00){\vector(1,0){59.67}}
\put(75.00,70.00){\vector(-1,0){50.00}}
\put(75.00,30.00){\vector(0,1){80.00}}
\put(120.00,30.00){\line(0,1){80.00}}
\put(120.00,110.00){\line(0,0){0.00}}
\put(39.67,30.00){\line(0,1){80.00}}
\put(99.33,85.67){\vector(-2,-1){0.2}}
\bezier{132}(113.00,110.00)(112.67,92.00)(99.33,85.67)
\bezier{112}(99.00,85.33)(85.67,73.67)(75.00,72.33)
\put(49.67,61.00){\vector(-1,-3){0.2}}
\bezier{140}(75.00,72.33)(53.67,74.00)(49.67,61.00)
\bezier{112}(49.33,60.67)(41.33,42.67)(42.67,34.00)
\put(52.33,86.00){\vector(4,-3){0.2}}
\bezier{116}(44.33,109.00)(41.67,93.00)(52.33,86.00)
\bezier{112}(52.33,86.00)(59.00,76.67)(75.00,72.67)
\put(96.67,60.33){\vector(1,-4){0.2}}
\bezier{128}(75.00,72.67)(94.33,72.67)(96.67,60.33)
\bezier{144}(96.67,60.33)(110.67,49.33)(114.33,31.67)
\put(75.00,25.33){\makebox(0,0)[cc]{r=0}}
\put(120.00,25.00){\makebox(0,0)[cc]{$r=r_{-}$}}
\put(39.67,25.00){\makebox(0,0)[cc]{$r=r_{-}$}}
\put(140.33,70.00){\makebox(0,0)[cc]{r}}
\put(19.67,70.00){\makebox(0,0)[cc]{r}}
\put(75.00,114.00){\makebox(0,0)[cc]{t}}
\put(90.00,87.33){\makebox(0,0)[cc]{$\gamma_{1}$}}
\put(60.00,87.33){\makebox(0,0)[cc]{$\gamma_{2}$}}
\put(107.33,59.67){\makebox(0,0)[cc]{$\gamma_{2}$}}
\put(50.67,48.67){\makebox(0,0)[cc]{$\gamma_{1}$}}
\put(75.00,15.00){\makebox(0,0)[cc]{Fig. (4)}}
\put(89.67,110.00){\makebox(0,0)[cc]{$III_{1}$}}
\put(57.67,110.33){\makebox(0,0)[cc]{$III_{2}$}}
\end{picture}
\section{Tidal Forces Near the Singularities}\label{sec5}
   A physical meaning of Riemann curvature tensor is that its components
describe tidal forces (relative accelerations) between two
particles in free fall. Consider a body falling towards a black
hole, then the effect of the gravitational attraction becomes
infinite as the singularity is reached. Jacobi vector fields
provide the connection between the behavior of nearby particles
and curvature, via the equation of geodesic deviation (Jacobi
equation)
\begin{equation}
\frac{D^{2}\eta^{a}}{D\tau^2} + R^{a}_{bcd}v^{b}v^{c}\eta^{d} = 0,
\end{equation}
where $v^a$ are the components of the tangent vector to geodesic
and  $\eta^a$ are the components of the connecting vector between
two neighboring geodesics.
\par
   In order to investigate in detail the behavior of Jacobi fields
we consider a congruence of timelike geodesics (path of particles)
with timelike unit tangent vector $v$ ($g(v,v) = - 1$). We define
at some point $q$ on the geodesic $\gamma(\tau)$ dual bases
$e_{1}^a, e_{2}^a, e_{3}^a, e_{4}^a$  and $e_{a}^1, e_{a}^2,
e_{a}^3, e_{a}^4$  of the tangent space $T_{q}M$ and dual tangent
$T^{\star}_{q}M$ respectively in the following way \cite{3} : We
choose $e_{4}^a$ to be $v^a$ and $e_{1}^a, e_{2}^a, e_{3}^a$ as
unit spacelike vectors, orthogonal to each other and to $v^a$. If
we parallelly propagate the basis along the timelike geodesic
$\gamma(\tau)$  (that is, $\frac{D}{D\tau}e^{a}_{i} = 0, \, i = 1,
2, 3$), $e^{a}_{4}$ will remain equal $v^a$, and $e_{1}^{a},
e^{a}_{2}, e^{a}_{3}$  will remain to orthogonal to $v^a$ (see
\cite{10} p. 80). The frame $e_{1}^a, e_{2}^a, e_{3}^a, e_{4}^a$
is called "parallel transported" (PT) frame. The orthogonal
connecting vector, $\eta^a$, between two neighboring timelike
geodesics may be expressed as $\eta^a  = \eta^{\alpha}
e_{\alpha}^a \, (\eta^4 = e_{\alpha}^{4}\eta^{\alpha} = 0)$, where
Greek indices take the value $1, 2, 3$.
\par
The Jacobi vector fields $\eta^a$ satisfy the following equation
\begin{equation}\label{4.2}
\frac{D\eta^{\alpha}}{D\tau} = \eta^{\alpha}_{;a}v^{a},
\end{equation}
\begin{equation}\label{4.3}
\frac{D^{2}\eta^{\alpha}}{D\tau^2} +
\tilde{R}^{a}_{bdc}e^{\alpha}_{a}v^{b}v^{c}e^{d}_{\beta}\eta^{\beta}
= 0,
\end{equation}
where $\eta^{\alpha}$, $\alpha = 1, 2, 3$, are the space-like
components of the orthogonal connecting vector $\eta^{a}$
connecting two neighboring particles in free fall; $\eta^4 = 0$.
The tilde denotes components in the PT frame and the components of
the Riemann tensor $\tilde{R}^{a}_{bdc}$ are given by
\begin{equation}\label{4.4}
\tilde{R}^{a}_{bdc} =
e_{e}^{a}e_{b}^{f}e_{c}^{g}e_{d}^{h}R^{e}_{fgh}.
\end{equation}
From (\ref{1.1}) the frame $e^{a}_{i}$ in Reissner-Nordstr\"{o}m
metric is given by :
\begin{equation}\label{4.5}
e_{1}^{a} =(1 - \frac{2m}{r} +
\frac{e^2}{r^2})^{\frac{1}{2}}(1,0,0,0),\\
\end{equation}
\begin{equation}\label{4.6}
e^{a}_{2} = r^{-1} (0, 1, 0, 0)
\end{equation}
\begin{equation}\label{4.7}
e^{a}_{3} = (r\sin\theta)^{-1}(0, 0, 1, 0)\\
\end{equation}
\begin{equation}\label{4.8}
e_{4}^{a} =
(1-\frac{2m}{r}+\frac{e^2}{r^2})^{-\frac{1}{2}}(0,0,0,1)\\
\end{equation}
The components of $\eta^{\alpha}$ can be written as follows
$$
\eta^{\alpha} = (\eta^{1}, \eta^{2}, \eta^{3}) = (\eta^{r},
\eta^{\theta}, \eta^{\phi}).
$$
Using (\ref{4.4}) - (\ref{4.8}),  $v^{a} = e^{a}_{4}$ and the
components of Riemann tensor for the metric (\ref{1.1}) (see
appendix), in (\ref{4.3}), we get
\begin{equation}\label{4.9}
\frac{D^{2}\eta^r}{D\tau^2}  = - \frac{3e^2 - 2mr}{r^4}\eta^r,\\
\end{equation}
\begin{equation}\label{4.10}
\frac{D^{2}\eta^{\theta}}{D\tau^2} =  \frac{e^2 - mr}{r^4}\eta^{\theta},\\
\end{equation}
\begin{equation}\label{4.11}
\frac{D^{2}\eta^{\phi}}{D\tau^2} =  \frac{e^2 - mr}{r^4}\eta^{\phi},\\
\end{equation}
In order to write equation (\ref{4.9}) - (\ref{4.11}) in terms of
ordinary derivative, we must evaluate the second covariant
derivative derivative
$\frac{D^2}{D\tau^2}$. Using $e^{a}_{4} = v^{a}$, equation(\ref{4.2}) takes the form:\\
\begin{equation}\label{4.12}
\frac{D\eta^{\alpha}}{D\tau} = \frac{d\eta^{\alpha}}{d\tau} +
\tilde{\Gamma}^{\alpha}_{ab}\eta^{b}v^{a},
\end{equation}
where
$$
\tilde{\Gamma}^{\alpha}_{ab} = e_{e}^{\alpha}e_{a}^{f}e_{b}^{g}\Gamma^{e}_{fg}.\\
$$ Differentiating (\ref{4.12}) covariantly,
using the Christoffel components of metric (\ref{1.1}) (see
appendix), we can write (\ref{4.9}) - (\ref{4.11}) in the form
\begin{equation}\label{4.13}
\frac{d^{2}\eta^r}{d\tau^2} = - \frac{3e^2 - 2mr}{r^4}\eta^r,\\
\end{equation}
\begin{equation}\label{4.14}
\frac{d^{2}\eta^{\theta}}{d\tau^2} =  \frac{e^2 - mr}{r^4}\eta^{\theta},\\
\end{equation}
\begin{equation}\label{4.15}
\frac{d^{2}\eta^{\phi}}{d\tau^2} =  \frac{e^2 - mr}{r^4}\eta^{\phi},\\
\end{equation}
If the quantity $3e^2 - 2mr$ is positive (negative), then this
indicates a tension or stretching (pressure or compression) in
radial direction. Similarly, if $e^2 - mr$ is positive (negative),
this indicates a pressure or compression (tension or stretching)
in the transversal directions.\\
\section*{Conclusion} The theory of differential spaces is used in
gluing two Reissner-Nordstr\"{o}m space-times $M_{1}$ and $M_{2}$
at their singularities .It is found that a light ray in the black
hole of $M_{1}$ can go continuously through the black hole of
$M_{2}$ . The behavior of nearby timelike geodesics with zero
angular momentum are investigated via the geodesic deviation
equations. It is shown that there are a tension (or stretching)
both in the radial and transversal direction according as the
quantities $3e^2 - 2mr$ and $e^2 - mr$ are positive(or
negative)respectively.
\section*{Appendix}
We use $(x^1,x^2,x^3,x^4)= (r, \theta, \phi, t)$ so that the
non-vanishing Christoffel symbols of the second kind of the line
element (1.1)are
$$
\Gamma^{1}_{11} = \frac{e^2 - mr}{r(r^2 - 2mr + e^2)},  \Gamma^{2}_{12}  = \frac{1}{r},\\
$$
$$
\Gamma^{1}_{22}  = - \frac{r^2 - 2mr +e^2}{r},     \Gamma^{3}_{13} =\frac{1}{r},  \\
$$
$$
\Gamma^{1}_{33}  = - \frac{(r^2 - 2mr + e^2)\sin^2\theta}{r}, \Gamma^{2}_{33}  =  - \sin\theta\cos\theta,  \\
$$
$$
\Gamma^{1}_{44}  =  \frac{(r^2 - 2mr + e^2)(\frac{m}{r^2} - \frac{e^2}{r^3})}{r^2},  \Gamma^{3}_{23} =  \cot\theta,  \\
$$
$$
\Gamma^{4}_{14}  = - \frac{r^2(-\frac{m}{r^2} +
\frac{e^2}{r^3})}{r^2 - 2mr + e^2}. \\
$$
\newpage
The non-zero Riemann tensor are:
$$
R^{1}_{212}  =  \frac{e^2 - mr}{r^2)},  R^{1}_{313}  =  \frac{(e^2 - mr)\sin^2\theta}{r^2},  \\
$$
$$
R^{1}_{414}  =  \frac{(r^2 - 2mr +e^2)(3e^2 -2mr)}{r^6},  R^{2}_{323}  = \frac{(2mr - e^2)\sin^2\theta}{r^2},  \\
$$
$$
R^{2}_{424}  = - \frac{(r^2 - 2mr + e^2)(e^2 - mr)}{r^6},
$$
$$
R^{3}_{434}  =  - \frac{(r^2 - 2mr + e^2)(e^2 - mr)}{r^6}.
$$

\end{document}